\documentclass{sig-alternate-10pt}

\pdfoutput=1

\usepackage[T1]{fontenc}
\usepackage[utf8]{inputenc}

\usepackage{graphicx}

\usepackage{url}

\usepackage{amsmath}
\usepackage{amssymb}

\usepackage{algpseudocode}
\algrenewcommand\algorithmicindent{1.0em}%

\usepackage[tight,footnotesize]{subfigure}

\usepackage{url}

\begin{document}

\title{A Flat and Scalable Data Center Network Topology Based on De Bruijn Graphs\titlenote{Technical Report TR-2016-05}}

\numberofauthors{1}

\author{
\alignauthor
Frank D\"urr\\
\affaddr{University of Stuttgart}\\
\affaddr{Institute of Parallel and Distributed Systems (IPVS)}\\
\affaddr{Universit\"atsstra\ss{}e 38}\\
\affaddr{70569 Stuttgart}\\
\email{frank.duerr@ipvs.uni-stuttgart.de}
}

\maketitle

\begin{abstract}
Due to the requirement of hosting tens of thousands of hosts in today's data centers, data center networks strive for scalability and high throughput on the one hand. On the other hand, the cost for networking hardware should be minimized. Consequently, the number and complexity (e.g. TCAM size) of switches has to be minimized. These requirements led to network topologies like Clos and Leaf-Spine networks only requiring a shallow hierarchy of switches---two levels for Leaf-Spine networks. The drawback of these topologies is that switches at higher levels like Spine switches need a high port density and, thus, are expensive and limit the scalability of the network.

In this paper, we propose a data center network topology based on De Bruijn graphs completely avoiding a switch hierarchy and implementing a flat network topology of top-of-rack switches instead. This topology guarantees logarithmic (short) path length. We show that the required routing logic can be implemented by standard prefix matching operations ``in hardware'' (TCAM) allowing for using commodity switches without any modification. Moreover, forwarding requires only a very small number of forwarding table entries, saving costly and energy-intensive TCAM. 
\end{abstract}

% ACM style
%\category{CR number}{category}{sub-category}[optional subject descriptor]
%\terms{Design, Performance, Measurement}
\section{Introduction}
\label{sec:introduction}
\noindent Today's data centers host tens of thousands of machines, and technology trends like the development of power-efficient and densely integrated micro-servers will allow for increasing the number of servers per data center even further. Consequently, a scalable data center network for connecting such large numbers of machines is one of the obvious requirements. Scalability comprises different metrics which should not grow fast while the number of machines is increased. These metrics include the path length between machines, number of required switches, and cost of hardware, e.g., due to large TCAM size or high port density of switches.

\begin{figure}[tb]
  \centering
  \includegraphics[scale=0.5]{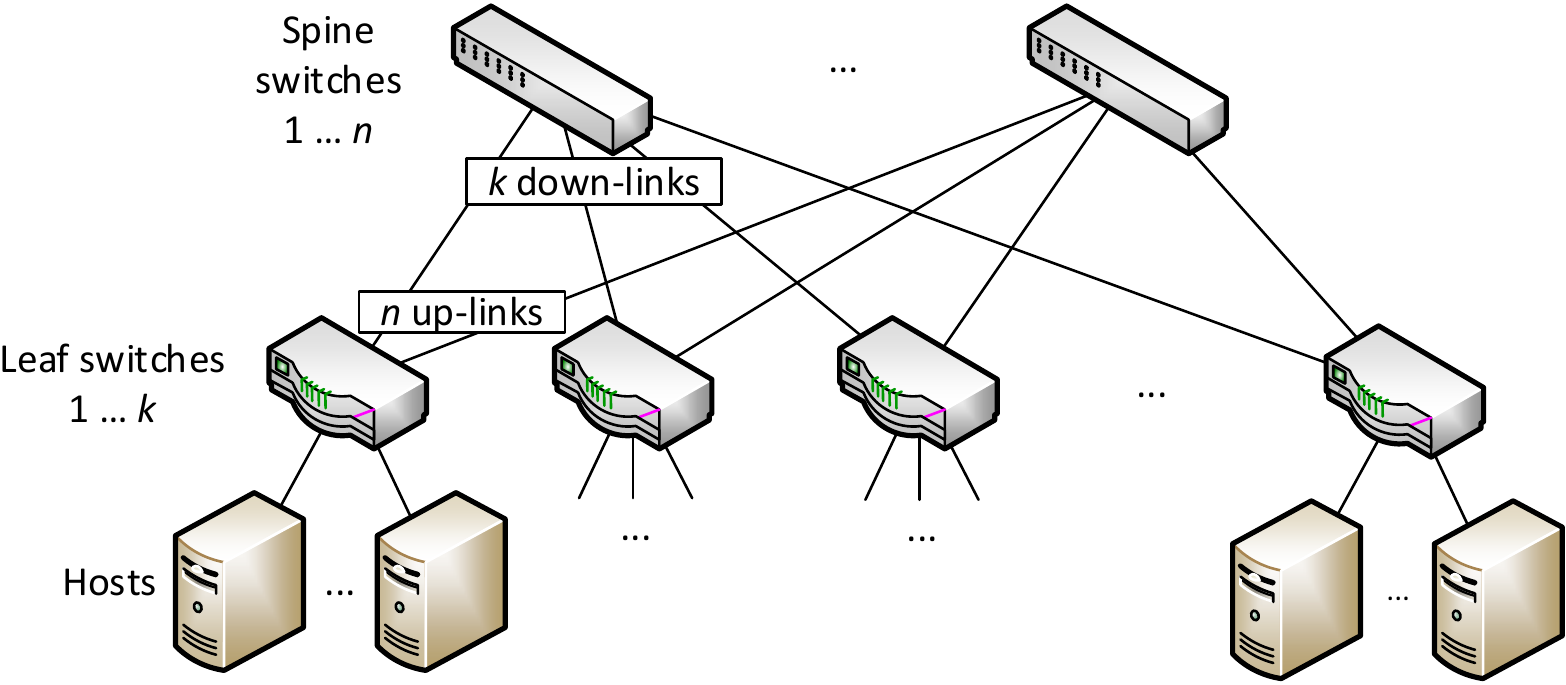}%
  \caption{Leaf-Spine topology}%
  \label{fig:leaf_spine_topology}%
\end{figure}

The most common data center network topology today are Clos networks \cite{Clo11}, in particular, folded three-stage Clos networks also called Leaf-Spine networks \cite{Cis10} (cf. Fig.~\ref{fig:leaf_spine_topology}). Here, we can see the trend to reduce the number of levels. The Leaf-Spine topology only contains two levels: the leaf switches with attached hosts, and the spine switches forming a multi-rooted hierarchical topology connecting hosts through a number of equally short paths (max. three switches between any pair of machines). Although this topology offers short paths and allows for load-balancing over several paths, using for instance ECMP, scaling it up obviously requires more powerful spine switches with high port density. Although switches with high port density exist, they are much more expensive than relatively simple top-of-rack-switches and also consume significant power. Therefore, it is a valid endeavor to search for alternative \emph{flat topologies} that do not require a spine level anymore and rather connect top-of-rack switches directly. Note that we do not strive for abandoning also top-of-rack switches and connecting servers directly since we think that switches are an efficient means to connect many commodity servers at high speed.

Since ToR switches typically offer a limited number of high-speed up-links (e.g., $4 \times 10\,\mathrm{Gbps}$ or $4 \times 40\,\mathrm{Gbps}$), constant-degree networks seem to be an ideal candidate for implementing a flat topology. Indeed, constant degree networks are well-known. For instance, 3D torus networks (6 links) have been used frequently in direct-connect high-performance networks \cite{ABC+05}. Moreover, in \cite{SHP+12} the authors have proposed to use constant-degree random graphs in the data center. Surprisingly, another very interesting graph structure has received much less attention as topology for data center networks: De Bruijn graphs \cite{dB46}. De Bruijn graphs have very interesting properties: (1) They are flat and have constant degree, so we can directly connect ToR switches through a constant number of up-links. (2) They theoretically scale better than $d$-dimensional torus networks (guaranteed logarithmic path length instead of $d$-th root). (3) Their structure implies a routing algorithm in contrast to random graphs---which also have logarithmic path length with high probability. In other words: random graphs have short paths, but it is harder to find them, whereas De Bruijn graphs have short paths which are easy to find by exploiting the De Bruijn graph structure.

Therefore, in this paper, we present concepts and mechanisms to implement a flat data center network topology of ToR switches based on De Bruijn graphs. In detail, we make the following contributions:

\begin{itemize}
\item We show that routing in a De Bruijn network can be implemented ``in hardware'' (TCAM) by using prefix matching operations only. Our implementation only requires a very small number of forwarding table entries, thus, saving energy-intensive and limited TCAM. Consequently, it can be implemented on inexpensive commodity switches.
\item We present mechanisms to implement the concept in a data center using an SDN-based architecture.
\item We compare the performance of De Bruijn routing to state-of-the-art ECMP routing in Leaf-Spine networks and to the performance of flat random graphs. Obviously, we cannot expect De  Bruijn networks to outperform Clos networks since even logarithmic paths are in general longer than three hops. Therefore, we want to answer the question, what is the price w.r.t. throughput decrease that we have to pay using flat De Bruijn networks on inexpensive switching hardware.
\end{itemize}

The remainder of this paper is structured as follows. First, we discuss related work in Section~\ref{sec:related_work}, before we present our system model in Section~\ref{sec:system_model}. In Section~\ref{sec:approach}, we present our approach to implement a De Bruijn data center network, whose performance is evaluated in Section~\ref{sec:evaluation}. Finally, we conclude our paper in Section~\ref{sec:summary} together with an outlook onto future work. 

\newpage % because of broken ACM template that does not keep section heading and first paragraph together ...
\section{Related Work}
\label{sec:related_work}
\noindent The demand for efficient data center networks has led to various network topologies. The most prominent topology is the already introduced Clos topology and its variant the Leaf-Spine network \cite{Cis10,MPF+09,ALV08,GHJ+09}. As discussed, this topology requires expensive hardware with high fan out at the top (Spine) level.

In particular in high performance computing, flat network topologies where servers are directly connected without switches are used. Torus networks \cite{ABC+05,ACR+10} are the prevalent flat routing structure here because of the ease of cabling (only neighboring hosts in the torus have to be connected) and the availability of multiple paths to increase the bandwidth. However, this comes at the price of longer paths ($\mathrm{O}(\sqrt[dim]{n})$ compared to $\mathrm{O}(\log(n)$). In \cite{PRI+10}, the authors use De Bruijn graphs for directly connected servers. In contrast, we use De Bruijn graphs between ToR switches. In particular, this brings up the question how to implement De Bruijn routing efficiently on existing switch hardware in TCAM.

Another approach is the application of random graphs to data center networks \cite{SHP+12}. From random graph theory, it is well known that such graphs have short logarithmic paths with $\log(n)$ links per node. However, since these are unstructured networks, the network includes no routing information making general routing algorithms necessary. Therefore, it is attractive to use \emph{structured networks} and exploit their structural information for routing. In \cite{SWS11}, the authors proposed to use the Kleinberg model, which achieves $\mathrm{O}(\log(n))$ path length with $\mathrm{O}(\log(n))$ links per node using greedy forwarding. In contrast, De Bruijn graphs achieve $\mathrm{O}(\log(n))$ path length with $\mathrm{O}(1)$ links.

Server-centric networks typically utilize cheap commodity switches and servers with multiple network interfaces as relay nodes. Approaches for container-size networks such as BCube \cite{GLH+09} are tailored to smaller networks of few thousand servers. In contrast, DCell \cite{GWT+08} defines a self-similar recursive structure with high scaling properties. We aim for similar scaling properties without requiring multi-port servers being involved in forwarding. 

\section{System Model}
\label{sec:system_model}
\noindent Before we present our approach, we first introduce our system model and assumptions. We assume a typical Cloud data center consisting of physical machines (hosts) hosting virtual machines (VMs). Each host is mounted in a rack and connected to the top-of-rack (ToR) switch of that rack through one network interface. Moreover, a virtual software switch (vSwitch) runs on each host. The vSwitch is connected to the VMs on the host and through the host's physical network interface to the ToR switch. 

Each ToR switch has a number of host links (one for each host in its rack), and a small number of up-links of higher bandwidth. A typical ToR configuration could be $48 \times 1\,\textrm{Gbps}$ host links and $4 \times 10\,\textrm{Gbps}$ up-links. As presented later, we use the up-links to link ToR switches to each other in a flat network of ToR switches. 

We assume that both ToR and vSwitches are multi-layer switches. Forwarding decisions between ToR switches are based on IP addresses. This assumption is crucial for our De Bruijn forwarding implementation since it relies on prefix matching. Between ToR and vSwitch and between vSwitch and host, we use MAC addresses of VMs for making forwarding decisions as presented below in detail.

Moreover, we use SDN to configure the flow tables of vSwitches and ToR switches by a logically centralized SDN controller with a global view onto the network. In particular, the SDN controller knows: (1) which VMs are located on which host; (2) the MAC and IP addresses of all hosts and VMs; (3) to which ToR switch and switch port the hosts are connected; (4) the De Bruijn network topology.

For a simpler description, we do not consider virtual LANs to isolate networks of different ``tenants'' or for splitting the data center network into sub-networks. However, our approach does not restrict the usage of these concepts in general.

\section{De Bruijn Data Center Network}
\label{sec:approach}
\noindent Next, we explain our De Bruijn network approach starting with an overview, and then explaining the De Bruijn topology and TCAM-based implementation in detail.

\subsection{Overview}
\label{sec:overview}
\noindent VMs are the sources and destinations of packets in our system. In order to give an overview, we consider the general case where the source VM ($\mathrm{VM}_\mathrm{src}$) and destination VM ($\mathrm{VM}_\mathrm{dst}$) are located in different racks, i.e., they are located on different hosts connected to different ToR switches. We split forwarding into three phases. \textbf{Phase~1:} Forwarding from $\mathrm{VM}_\mathrm{src}$ to the source ToR switch. \textbf{Phase~2:} Multi-hop De Bruijn forwarding from the source ToR switch to the destination ToR switch where the host of $\mathrm{VM}_\mathrm{dst}$ is connected. \textbf{Phase~3:} Forwarding from the destination ToR switch to $\mathrm{VM}_\mathrm{dst}$ via the vSwitch of the destination host.

Phase~2 uses special IP addresses derived from De Bruijn identifiers and prefix matching as presented in Sec.~\ref{sec:de_bruijn_forwarding}. Phase~3 uses the identity of $\mathrm{VM}_\mathrm{dst}$. Therefore, we distinguish between the \emph{location} (destination ToR switch) of the destination VM  for forwarding between racks and the VM's \emph{identity}. A similar concept is used, for instance, by the Location/Identifier Separation Protocol (LISP) \cite{RFC6830}. There are two alternatives to define the identity of $\mathrm{VM}_\mathrm{dst}$: $\mathrm{VM}_\mathrm{dst}$'s MAC address or its IP address. If we use IP addresses, we will have to encapsulate IP packets since Phase~2 uses different (locator) IP addresses for inter-ToR-switch forwarding. Since in a LAN setting we can use MAC addresses without the overhead of an additional encapsulation header, and because OpenFlow so far does not support the encapsulation of IP packets in IP packets, we decided to use the MAC addresses of VMs as their identifiers. 

Based on this decision, we can now describe the three phases in detail. Assume $\mathrm{VM}_\mathrm{src}$ wants to send a packet to $\mathrm{VM}_\mathrm{dst}$. In Phase~1, $\mathrm{VM}_\mathrm{src}$ first sends an ARP request via its local vSwitch, which is intercepted by the vSwitch and re-directed to the SDN controller. The SDN controller answers the request with the MAC address of $\mathrm{VM}_\mathrm{dst}$. $\mathrm{VM}_\mathrm{src}$ then forwards the packet to the local vSwitch. Assume that the vSwitch so far has no entry matching the IP address of $\mathrm{VM}_\mathrm{dst}$. It therefore re-directs the packet to the SDN controller. The SDN controller configures the vSwitch with a flow table entry (flow for short) that re-writes the outgoing packet's IP address to the De Bruijn IP address (locator address of the destination ToR switch), which is used in Phase~2. Note that the MAC address of $\mathrm{VM}_\mathrm{dst}$ stays valid and can be used to identify $\mathrm{VM}_\mathrm{dst}$ in later steps. After IP re-writing, the packet is forwarded to the ToR switch of $\mathrm{VM}_\mathrm{src}$, which starts Phase~2.

In Phase~2, ToR switches perform De Bruijn forwarding as described in Sec.~\ref{sec:de_bruijn_forwarding} using locator addresses derived from De Bruijn identifiers of ToR switches. A small number of flows called \emph{De Bruijn flows} in the following is required per ToR switch, which are static and independent of the locations of VMs. Therefore, the SDN controller can proactively configure these flows.

To implement the transition from Phase~2 to Phase~3, the SDN controller installs two types of flows with different priorities in ToR switches: The already mentioned De Bruijn flows (low priority) used during Phase~2; \emph{identifier flows} using VM MAC addresses as matching criterion (high priority) used during Phase~3. For each ToR switch, the SDN controller installs an identifier flow for each VM hosted on the local hosts of the ToR switch.
%
%\begin{displaymath}
%\mathrm{MAC}(\mathrm{VM}_\mathrm{dst}) \mapsto \mathrm{OUTPUT}(\mathrm{PORT}(\mathrm{VM}_\mathrm{dst}))
%\end{displaymath}
%
If a packet targeted at a local VM arrives at the ToR switch during Phase~2, the higher priority identifier flow is effective and forwards the packet to the local host where the VM is located. Similarly, vSwitches are configured by the SDN controller with identifier flows for each VM connected to the vSwitch. In addition, flows at vSwitches re-write the locator IP destination address to the IP address of $\mathrm{VM}_\mathrm{dst}$. These flows are installed at the same time when the vSwitch of $\mathrm{VM}_\mathrm{src}$ is configured in Phase~1.
%
%\begin{displaymath}
%\mathrm{mac}(\mathrm{VM}_\mathrm{dst}) \mapsto [\mathrm{setip}(\mathrm{ip}(\mathrm{VM}_\mathrm{dst})),\mathrm{out}(\mathrm{port}(\mathrm{VM}_\mathrm{dst}))]
%\end{displaymath}
%

Note that ToR switches only need a small number of De Bruijn flows as shown in Sec.~\ref{sec:de_bruijn_forwarding} and one entry for each local VM, independent of the total number of VMs in the data center. Moreover, only De Bruijn flows require IP prefix matching. The flow entries for VMs use exact matches on MAC addresses. Even inexpensive ToR switches allow for thousands of flows with matches on MAC addresses. vSwitches have one flow per local VM and one for each active destination VM, which easily can be managed by a software switch whose flow table size is not restricted by TCAM.

\subsection{De Bruijn Topology and Routing}
\label{sec:de_bruijn_forwarding}
\noindent To understand the topology of ToR switches and De Bruijn routing, we have to introduce De Bruijn graphs first. In a De Bruijn graph, each vertex has a label $L = (\lambda_1,\ldots,\lambda_m)$, which is a string of $m$ digits $(\lambda_1,\ldots,\lambda_m)$ with $\lambda_i \in \{0,\ldots,d-1\}$. The total number of vertices is defined as $n = d^m$. A vertex has one outgoing edge for each digit $\lambda_i \in \{0,\ldots,d-1\}$. The edge for digit $\lambda$ from the source vertex with label $L_\mathrm{src}$ points to the destination vertex with label $L_\mathrm{dst} = \mathrm{leftshift}(L_\mathrm{src}) \oplus \lambda$. i.e., $\lambda$ is shifted in from the right evicting the left-most digit.

\begin{figure}[tb]
  \centering
  \includegraphics[scale=0.6]{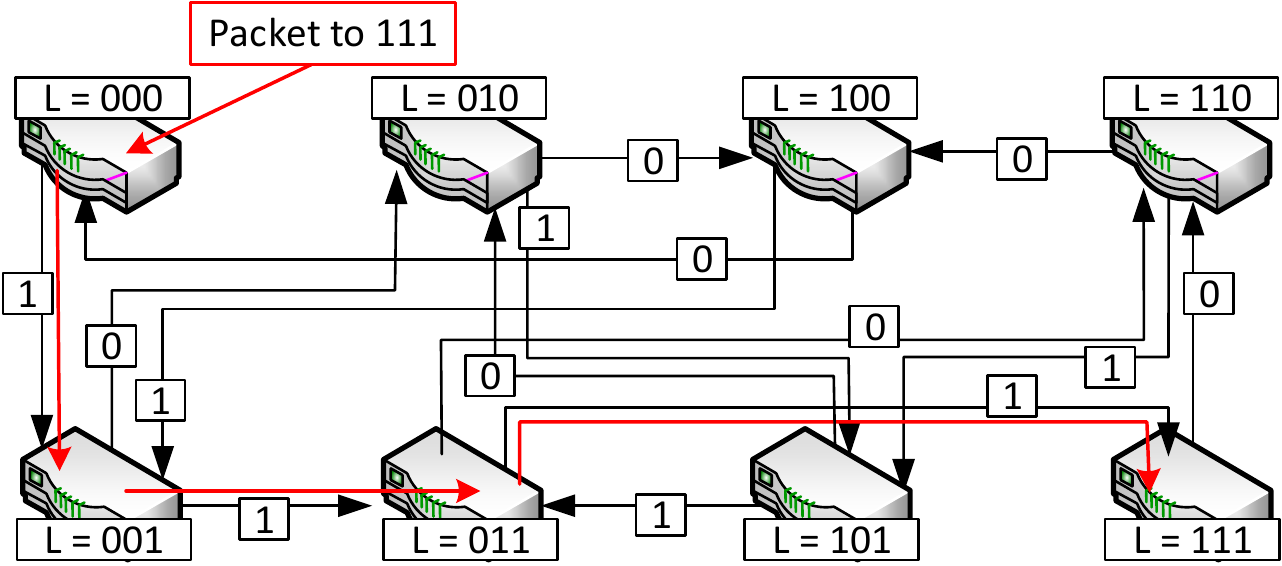}%
  \caption{De Bruijn Topology $d = 2$, $m = 3$}%
  \label{fig:debruijn_topology}%
\end{figure}

We now use De Bruijn graphs to define the topology of ToR switches (cf. Fig.~\ref{fig:debruijn_topology}). Each ToR switch corresponds to a vertex of the De Bruijn graph with label $L$. Edges become inter-ToR-switch links implemented by a maximum  number of $2d$ up-link ports, e.g., $4 \times 10\,\mathrm{Gbps}$ or $16 \times 10\,\mathrm{Gbps}$ (using break-out cables on four $40\,\mathrm{Gbps}$ up-links). Label length $m$ is chosen according to the desired number $n$ of switches such that $n = d^m$. 

De Bruijn graphs are directed graphs. However, network links are usually bi-directional in a switched network. Therefore, 50 \% of the bandwidth would be wasted if links were only used in one direction. We solve this problem by embedding two logical De Bruijn graphs using the same nodes with same labels and links but inverted and re-labeled edges (right-shift instead of left-shift of labels). For instance, in Fig.~\ref{fig:debruijn_topology}, we would add an edge from $110$ to $111$ with label $1$ ($1 \oplus \mathrm{rightshift}(110)$). Forwarding on these two graphs can be implemented using two sets of flow table entries marked by different prefixes. This concept allows for link failure handling and load balancing using alternative paths by selecting the prefix during the configuration of the source vSwitch in Phase~1 (see above: IP address re-writing to De Bruijn IP address). For a simpler description, we only consider one De Bruijn graph based on the left-shift operation in the following. 

Routing is based on the definition of edges in the De Bruijn graph. Let $L_\mathrm{dst}$ be the label of the ToR switch where $\mathrm{VM}_\mathrm{dst}$ is located, e.g., $111$ in Fig.~\ref{fig:debruijn_topology}. Conceptually, in each forwarding step at a ToR switch with label $L_\mathrm{switch}$ the switch calculates the longest sub-label $L'$ with $|L'| \leq m$ where $L'$ is a prefix of $L_\mathrm{dst}$ and a suffix of $L_\mathrm{switch}$. If $L' = L_\mathrm{dst} = L_\mathrm{switch}$, then the destination switch has been reached, which has a high priority flow based on the $\mathrm{VM}_\mathrm{dst}$'s MAC address. Otherwise, the current switch tries to extend the matching prefix/suffix $L'$ by one digit by forwarding the packet along the edge with digit $\lambda$, where $\lambda$ is equal to the $|L'|+1$-th digit of $L_\mathrm{dst}$. For instance, in Fig.~\ref{fig:debruijn_topology} the packet addressed to $L_\mathrm{dst} = 111$ is forwarded by switch $L_\mathrm{switch} = 001$ along edge $1$ to extend $L' = 1$ to $L'' = 11$. The path length grows as $O(m = \log_d(n))$ since we must at most shift in $|L_\mathrm{dst}| = m$ digits to completely match $L_\mathrm{dst}$.

To make this routing algorithm applicable to standard switches, we now show how to implement this algorithm by using flows with standard IP prefix matching. Figure~\ref{alg:debruijn_flow_config} shows the algorithm executed by the SDN controller to program the flow table of a ToR switch assigned to the De Bruijn vertex with label $L_\mathrm{switch}$. To find the neighbors of the switch, the SDN controller has an internal representation of the De Bruijn graph. The idea is now, to enumerate the suffixes of $L_\mathrm{switch}$ starting with the suffix of length zero (line~\ref{line:suffixlen}). For each De Bruijn edge, each suffix is extended by one digit $\lambda$ (line~\ref{line:extension}), which corresponds to the link to a neighboring switch in the De Bruijn topology that would extend $L'$ by this one digit after forwarding (line~\ref{line:neighbor}). The extended suffix $L'' = \mathrm{leftshift}(L') \oplus \lambda$ becomes the IP address prefix of the flow's match (after translating labels to binary IP addresses) with a subnet mask of corresponding length. Note that we do not generate a flow if the extended suffix is again a suffix of $L_\mathrm{switch}$ (line~\ref{line:checksuffix}) since $L'$ is defined as the \emph{longest} matching prefix/suffix of $L_\mathrm{dst}$ and $L_\mathrm{switch}$. 

\begin{figure}[tb]
\begin{algorithmic}[1]
\Procedure{ConfigureDeBruijnFlows}{$\mathrm{switch}$}
  \ForAll{$\mathrm{len} \in \{0,\ldots,m-1\}$} \label{line:suffixlen}
    \State $L' \gets \mathrm{suffix}(L_\mathrm{switch},\mathrm{len})$ \label{line:extension}
    \ForAll{$\lambda \in \{0,\ldots,d-1\}$}
       \State $L''  \gets \mathrm{leftshift}(L') \oplus \lambda$
       \If{$L''$ not suffix of $L_\mathrm{switch}$} \label{line:checksuffix}
         \State $\mathrm{port} \gets \mathrm{port}(\lambda \textrm{-edge of } \mathrm{vertex}(L_\mathrm{switch}))$ \label{line:neighbor}
         \State $f \gets (\mathrm{ip}(L''),\mathrm{mask}(L'')) \mapsto \mathrm{port})$
         \State addflow($\mathrm{switch}$, $\mathrm{flow}=f$, $\mathrm{priority}=\mathrm{len}$) \label{line:priority}
       \EndIf
    \EndFor
  \EndFor
\EndProcedure
\end{algorithmic}
\caption{Configuration of De Bruijn flows}
\label{alg:debruijn_flow_config}
\end{figure}

Table~\ref{tab:fwd_table} shows the De Bruijn flows of switch $101$ from Fig.~\ref{fig:debruijn_topology}. Note that prefixes might overlap. For instance, switch $101$ has flows with the prefixes $0\mathrm{**}$ and $011$. Therefore, we search for the \emph{longest} matching prefix by assigning a flow priority based on the prefix length (line~\ref{line:priority}).

\begin{table}
\begin{tabular}{l|l|l|l}
Label Prefix & IPv4/Mask & Priority & Out port \\
\hline
\verb|0**| & \verb|0.0.0.0/30| & 1 & $\mathrm{port}(\lambda = 0)$ \\
\verb|10*| & \verb|0.0.0.4/31| & 2 & $\mathrm{port}(\lambda = 0)$ \\
\verb|11*| & \verb|0.0.0.6/31| & 2 & $\mathrm{port}(\lambda = 1)$ \\
\verb|010| & \verb|0.0.0.2/32| & 3 & $\mathrm{port}(\lambda = 0)$ \\
\verb|011| & \verb|0.0.0.3/32| & 3 & $\mathrm{port}(\lambda = 1)$ \\
\end{tabular}
\caption{De Bruijn Flows of Switch $101$}
\label{tab:fwd_table}
\end{table}

As obvious from the loops of Algorithm~\ref{alg:debruijn_flow_config}, the number of De Bruijn flows per switch grows as $\mathrm{O}(2md)$ since the algorithm can create only one flow for each (suffix-length,digit)-combination, and we embed two De Bruijn graphs into the network. For instance, with four up-links per switch ($d = 2$) and $32,768 = 2^{15}$ switches ($m = 15$), we do not need more than $2 \times 30$ De Bruijn flow table entries per switch, while guaranteeing a maximum path length of 15 switches. Increasing the number of up-links to 16 decreases the maximum path length to 5 and increases the maximum flow table size to $2 \times 40$. 

\section{Evaluation}
\label{sec:evaluation}
\noindent Finally, we compare the performance of our approach to Leaf-Spine networks and another flat constant-degree network based on random graphs similar to \cite{SHP+12}. 
%In our comparison to Leaf-Spine networks, we want to find out which performance penalty we have to pay for abandoning the Spine layer and using a flat topology instead. The comparison to random graphs, which are also known to have logarithmic path length, shows whether the light-weight De Bruijn routing algorithm with its small number of static flows is competitive to common shortest path routing, which requires one entry per destination ToR switch.

\subsection{Evaluation Setup}
\noindent Although we also did experiments in emulated networks with Mininet, the following results are based on network simulations with OMNeT++ to also evaluate larger topologies. We use ToR switches with 40 hosts per switch connected through $1\,\mathrm{Gbps}$ links and $4 \times 10\,\mathrm{Gbps}$ up-links to four Spine switches (Leaf-Spine topology) or other ToR switches (De Bruijn and random topology). Consequently, we use De Bruijn topologies based on binary digits ($d = 2$) and four Spine switches in the Leaf-Spine topology. The constant-degree random topology adds four links between randomly selected ToR switches. Switches use tail-drop queues with a queue size of 4096 entries.

As performance metric, we use the throughput of TCP connections. We did not simulate VMs but placed TCP processes directly on hosts. On each host, we placed one TCP sender, which sends at highest possible rate to a TCP receiver on a randomly selected other host.

We compared several topologies and routing algorithms: \emph{LS/ECMP}: Equal Cost Multipath Routing (ECMP) on the Leaf-Spine topology. ECMP performs load balancing by distributing TCP connections across all shortest paths. \emph{Random/ECMP}: ECMP routing on the random topology. \emph{DB/DBRouting}: De Bruijn topology and De Bruijn routing using the shorter path of the two embedded De Bruijn graphs. \emph{DB/ECMP}: ECMP routing on the De Bruijn topology, i.e., not using the De Bruijn routing algorithm, to evaluate the quality of the topology and routing algorithm separately.

\subsection{Results}
\noindent In the following experiments we varied the number of ToR switches from 8 to 128 (320 to 5120 hosts). First, we let all TCP connections ``warm-up'' and then calculate the average throughput of all TCP connections in a time window of $0.5\,\mathrm{s}$. Each experiment consists of 5 simulation runs with different random connection distributions and random topologies for Random/ECMP.

\begin{figure}[tb]
  \centering
  \includegraphics[scale=0.9]{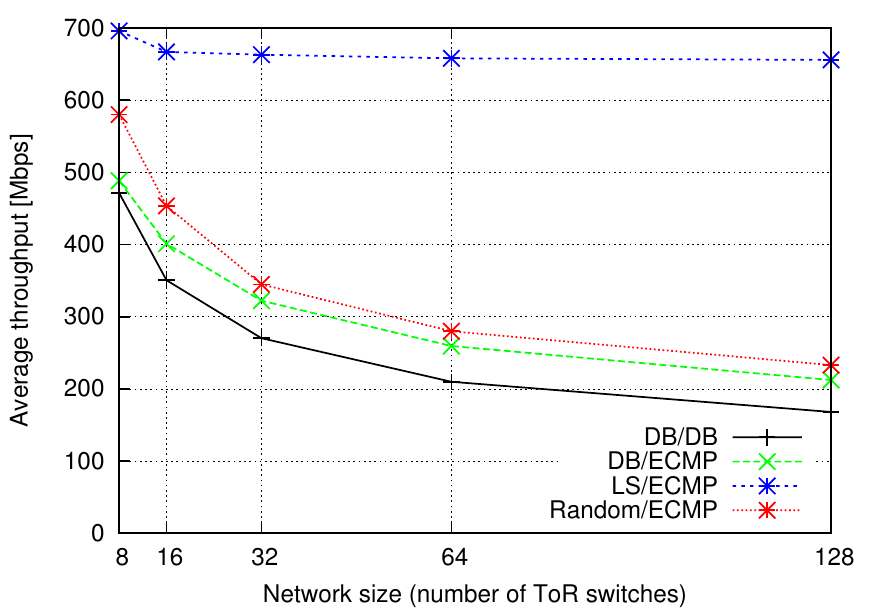}%
  \caption{Throughput}%
  \label{fig:throughput_40x1_4x10}%
\end{figure}

Figure~\ref{fig:throughput_40x1_4x10} shows the throughput over the network size. As expected, LS/ECMP has the highest performance since the path length stays constantly small (max. 3 hops) while scaling the network. In contrast, the performance of both flat networks, which both have $\mathrm{O}(\log{n})$ path length, decreases with increasing switch number: Compared to LS/ECMP, the throughput of DB/DBRouting decreases from 67 \% for 8 switches down to 25 \% for 128 switches; Random/ECMP achieves 83 \% down to 35 \% of the throughput of LS/ECMP. Comparing the two flat topologies, DB/DBRouting has between 81 \% and 72 \% of the performance of Random/ECMP. Note that the superior performance of LS/ECMP comes at the price of increasing the port density of Spine switches from $8 \times 10\,\mathrm{Gbps}$ to $128 \times 10\,\mathrm{Gbps}$ (and so forth for larger topologies). It is also important to note that in a real data center, the performance of DB and Random might be even closer to LS since we have (deliberately) chosen a challenging scenario where TCP connections are randomly distributed between hosts, which are with high probability located in different racks (88 \% to 99 \% inter-rack traffic for the 8 and 128 switch topology, respectively). As shown in \cite{BAM10}, in cloud data centers, 80 \% of the traffic stays \emph{within} the rack.

We also see that the throughput of DB/DBRouting is between 97 \% and 79 \% of the throughput of DB/ECMP. Thus, we can state that using De Bruijn routing with its very small number of static flows (max. 28 De Bruijn flows per ToR switch for the 128 switch/5120 hosts topology) is close to the performance of the more complex ECMP routing algorithm, and the limiting factor compared to LS/ECMP seems to be mostly the De Bruijn topology rather than the routing algorithm.

\textbf{Takeaway messages:} 
(1) Leaf-Spine/Clos networks have significantly higher throughput than flat constant-degree networks, but require more expensive switches with high port density to scale. (2) Flat De Bruijn networks have about 25 \% less throughput compared to flat constant-degree random graphs, however, without requiring relatively complex ECMP routing---De Bruijn routing can be directly implemented ``in hardware'' (TCAM) and requires only very small TCAM space.

\section{Summary and Future Work}
\label{sec:summary}
\noindent In this paper, we presented a data center network topology based on De Bruijn graphs. On the one hand, this flat topology does not require powerful switches with high port density to scale. Moreover, the De Bruijn graph structure allows for implementing a routing algorithm with guaranteed logarithmic path length ``in hardware'' (TCAM) with very small TCAM size requirements. On the other hand, our evaluations show that there is a price to pay w.r.t. performance. If there is a high ratio of inter-rack traffic, Leaf-Spine networks can achieve significantly higher performance than the flat De Bruijn topology, so the (monetary) cost for high-port-density Spine switches might be well-spent in such scenarios. Moreover, we showed that within the class of flat constant-degree networks, De Bruijn networks are very competitive.

An interesting question for future research is, how to connect a flat data center network like a De Bruijn network or random graph topology to the Internet via gateways, e.g., where and how many gateways should be placed within the flat topology? 

\bibliographystyle{abbrv}
\bibliography{bibliography}

\end{document}